# Taming Atomic Defects for Quantum Functions

S. M. Hus, A.-P. Li

Center for Nanophase Material Sciences, Physical Sciences Directorate, hussm@ornl.gov, apli@ornl.gov

## INTRODUCTION

Alice needed to go through a rabbit hole to reach Wonderland. But for us, the best way to reach the quantum wonderland [1] may mean never leaving it in the first place. All materials, regardless of how trivial they are, are governed by the rules of quantum mechanics at the atomic level. However, in most cases, these fundamental properties are diluted into a classical picture in the macroscopic world. To harness these hidden properties, we usually look at so-called "quantum materials," which expose the quantum mechanical nature of matter to our view at macroscopic scales. At this scale, the quantum functions show up in various forms such as superconductivity or nontrivial electronic band topology. On the other hand, by simply staying in the atomic scale we may be able to tame more fundamental quantum functions like entanglement and single-photon emission.

Single atoms provide an ideal system for utilizing fundamental quantum functions. Their electrons have well-defined energy levels and spin properties. Even more importantly, for a given isotope—say, $^{12}$C—all the atoms are identical. This creates a perfect uniformity that is impossible to achieve in macroscopic-size quantum systems. However, herding individual atoms is a very difficult task that requires trapping them with magnetic or optical means and cooling them down to temperatures in the nanokelvin range [2]. On the other hand, the counterpart of single atoms—the single defects—may be as good as atom-based quantum systems if not better. These defects, also referred as quantum defects, possess the favorable energy, spin, and uniformity properties of single atoms and remain in their place without the help of precisely tuned lasers. While the number of usable isotopes is set, the combinations of defects and their host material are practically limitless, giving us the flexibility to create precisely designed and controlled quantum systems. Furthermore, as we tame these defects for the quantum world, we bring about transformative opportunities to the classical world in forms such as ultradense electronic devices and precise manufacturing.

In this research insight, we introduce some of our recent work on precisely controlled creation and manipulation of individual defects with a scanning tunneling microscope (STM). We also discuss possible pathways for utilizing these capabilities for the development of novel systems for Quantum Information Science (QIS) applications such as quantum information processing and ultrasensitive sensors.

## BACKGROUND

Atomic-scale defects such as vacancies and substitutions (i.e., dopants) play an important role in controlling the electronic properties of materials. Today, most electronic devices rely on the sheer concentration of defects rather than the individual defects themselves. Utilizing single defects for classical or QIS applications faces two challenges. First, the defects are usually buried deep in the bulk, limiting access to the defect and making it hard to understand the exact defect structure. Second, creating the desired defect structures requires the capability to manipulate individual atoms with atomic precision.

The first challenge has been mostly addressed with the advent of 2D materials over the last two decades. These materials bring every single atom and defect within the range of optical and electronic probes. Furthermore, some of the most popular 2D materials—hexagonal boron nitride (hBN) [3] and transition metal dichalcogenides (TMDs) [4], for example—have already shown us that they have defect structures that act as single-photon emitters (SPEs) [5] at room temperature. However, 2D materials host many native defects that come in different variations and are randomly distributed in the 2D material. Therefore, despite the tremendous efforts to identify the exact structure of SPEs, it is still not clear which defect(s) are responsible.

While being equally important, the second challenge has been tackled far less than the first one. The randomness in defect distribution takes us back to today's electronic devices, for which only the defect concentration for large-scale devices can be safely counted on. If the defects are to be used individually in ultradense electronic devices or as qubits in QIS applications, their precise structure, location, spatial separation, and environment should be observed and controlled with atomic resolution. STM is one of the few techniques that can address this challenge, as it is equipped with advanced imaging, spectroscopy, and manipulation capabilities, all at the individual atom level. These capabilities make it possible to identify individual defects and create or manipulate the desired defect structures by moving atoms into a chosen geometry [6].

Our current research aims to address both challenges. We use STM to identify host material and defect structures, link structures with the desired functionalities, and develop techniques to individually modify defects to gain new functions as illustrated in Figure 1.

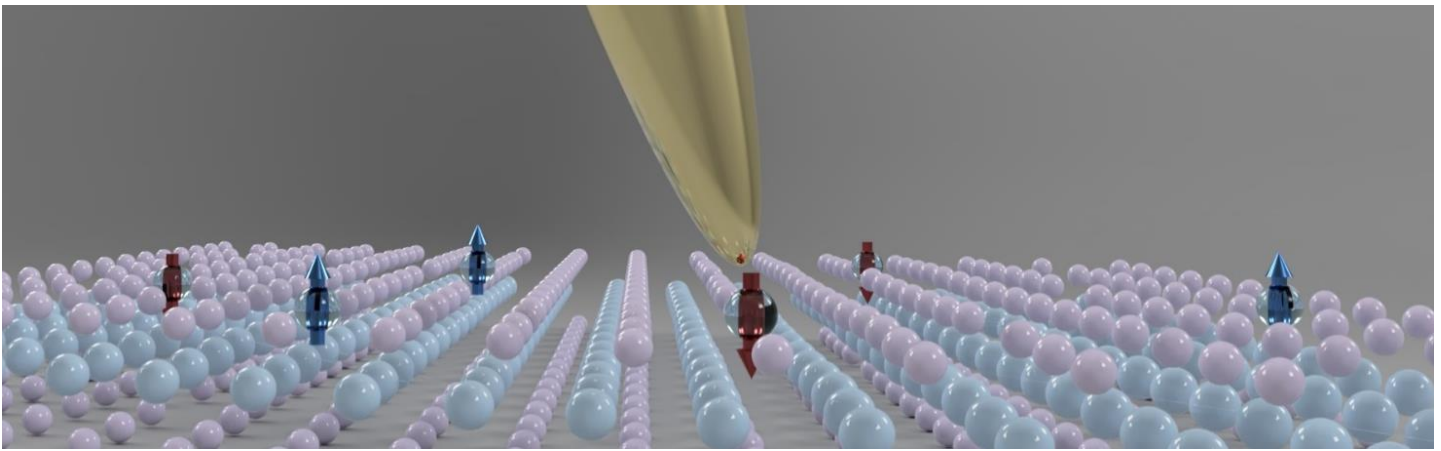

Figure 1. Scanning tunneling microscopy (STM) builds on the fundamental principle of quantum tunneling, allows imaging, spectroscopy, and manipulation, all at the individual atom level, and offers the capability of taming individual defects for deterministic quantum functions. For instance, STM can be used to manipulate the spin orientation (red and blue arrows) and location of individual defects (clear balls) in the crystal lattice of 2D materials (pink and blue balls)  (Source: Saban Hus)

*This manuscript has been authored by UT-Battelle LLC under contract DE-AC05-00OR22725 with the US Department of Energy (DOE). The US government retains and the publisher, by accepting the article for publication, acknowledges that the US government retains a nonexclusive, paid-up, irrevocable, worldwide license to publish or reproduce the published form of this manuscript, or allow others to do so, for US government purposes. DOE will provide public access to these results of federally sponsored research in accordance with the DOE Public Access Plan (http://energy.gov/downloads/doe-public-access-plan).*

## RESULTS

**Development of single-defect memory devices.** Since the development of the first integrated circuits, the size of basic electronic devices (e.g., transistors, resistors, memory units) has been scaling down. In this race to miniaturization, the ultimate goal has been reaching a scale where a single atom controls the device's function. We achieved this goal by creating a two-terminal memory unit, also known as memristor, which can switch between 0 and 1 states by moving a single gold atom in and out of a vacancy defect in atomically thin TMD layers [7]. These memory devices are composed of monolayer 2D materials sandwiched between two gold electrodes [8]. In this work, we first identified and characterized the defects of the 2D material with STM. Then we positioned a gold STM tip at the individual defect locations and used it in contact mode to act as the top electrode of the nanoscale device. We demonstrated that a single chalcogenide vacancy constitutes the active region of the device. Under a proper electric field created by the STM, a gold atom from the electrodes can be moved in and out of this vacancy, which reversibly tunes the local electronic structure between metallic and insulating characteristics, respectively (Figure 2). In this work, we demonstrated that quantum defects can be modified with atomic precision to create the smallest classical memory devices.

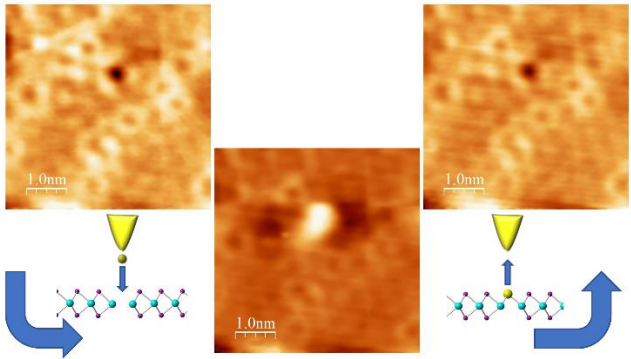

Figure 2. Individual vacancy defects in TMD monolayers can be filled with gold atoms with field-induced atom migration from the tip. This reversible process alters local electronic states between insulating and metallic characteristics, creating binary states of atomic size memories. The method is applicable to many other quantum defect–metal atom pairs and can be used to create designed defect structures. (Source: Hus et al.)

The ongoing pursuit is utilizing these skills to create and test novel magnetic defect structures. Multiple theoretical studies indicate that the defects in 2D materials can host spin-polarized electronic states due to strong spin-orbit coupling [9]. Experimentally, STM studies have already confirmed the existence of a large ~250 meV splitting between spin-polarized defect states of sulfur vacancies in $WS_2$ [10]. Filling these vacancies with different magnetic or nonmagnetic atoms may create even more pronounced and tunable spin states. The functionality of these defects is similar to spin-polarized magnetic adatoms on metal surfaces [10], which are seen as possible candidates for qubits and spintronic applications. But magnetic adatoms are extremely mobile on metal surfaces and require very low temperatures to stabilize and manipulate. On the other hand, most defects in 2D materials are spatially and energetically stable at room temperature, making them more amenable to experiments and applications.

STM manipulation allows us to go beyond randomly distributed defects and create deterministic ones. For instance, defect structures with longer spin coherence time can be designed, built, and tested. Similarly, defect pairs with a controlled spatial separation can be created to observe the interaction between their spin states. Furthermore, our multiprobe STM capabilities [11] enable us to tune this interaction by using one of the probes as a gate between two defects.

**3D defect engineering.** Defect manipulation in 2D materials is also possible beyond monolayer films. Using STM, we can identify the 3D lattice locations of selenium vacancy ($V_{Se}$) defects near the surface of layered pentagonal TMD-$PdSe_2$ [12]. Furthermore, we can reversibly switch the defects between neutral and negatively charged states and trigger their migration within the lattice, by using STM both as a movable electrode and characterization probe.

In STM, a small bias voltage (typically <2V) is applied between tip and sample to create and measure the tunneling current between them. When the STM tip is very close to the surface, this bias acts as a gate voltage and locally bends the electronic bands of the semiconductor materials. If the semiconductor has a defect state within the band gap and close to the Fermi level, this band bending may push the defect states across the Fermi level. As a result, the defects can be switched between neutral and charged states. Crossing the Fermi level creates a step in the tunneling current that is displayed as charging rings in dI/dV maps. The band-bending effect is reduced as we get farther away from the tip. Therefore, the defects at the lower layers display smaller charging rings. By measuring the size of these rings, we can identify the depth of the defect with high precision (Figure 3).

Compared to hexagonal TMDs, the unique pentagonal lattice of $PdSe_2$ has weaker covalent bonds and holds lower barrier energies for vacancy migration. This makes it possible to both "write" and "erase" an atomic defect from a particular lattice site by applying slightly higher voltage pulses (~2.0 V) than typical bias (Figure 3). Once the desired vacancy geometry is created, the vacancies at the top layer can be filled with substitute atoms to add new functionalities. Furthermore, the vacancies in the underlying layers can be utilized to modify the local charge and energy landscape.

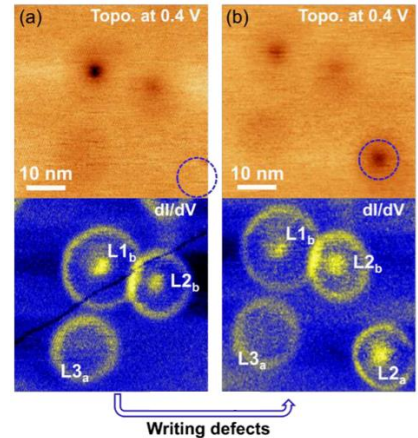

Figure 3. Individual vacancy defects can be written and erased (not shown) by applying voltage pulses to the surface of a layered material, $PdSe_2$. The size of charging rings around the defects can be used to identify the depth of the defect. (Source: Nguyen et al.)

**Magnetic end states in atomically precise graphene nanoribbons.** Graphene is one of the original quantum materials in which the electrons behave as massless Dirac fermions. In its pristine 2D form, graphene does not have a bandgap, which limits its direct utilization in QIS applications. However, lattice defects, including the edges, break graphene's structural symmetry and create a local environment where the electronic structure is significantly modified. In the 1D form of planer graphene, namely graphene nanoribbons (GNRs), an electronic bandgap is opened due to quantum confinement, and nontrivial electronic states can emerge. In their atomically precise form, these nanoribbons are of great interest for QIS applications due to their spin-polarized edge states; coupled spin centers; and highly tunable electronic, optical, and transport properties [13].

GNRs are typically synthesized with metal surface–assisted chemical reactions, leaving us with nanoribbons on a noble metal substrate. However, the metallic surface states of the substrate strongly couple with the electronic states of the GNR and suppress its distinctive properties. Transferring the nanoribbons to insulating substrates is also challenging because any alteration in its atomically precise edges will modify the electronic states of interest. Therefore, using a nonmetallic substrate for synthesis is highly desirable. Using specifically designed precursor molecules, we synthesized precise GNRs on the surface of rutile titanium dioxide ($TiO_2$), which is a large bandgap insulator [14]. Our STM studies on these nanoribbons confirmed the formation of planar armchair GNRs with well-defined zigzag ends as well as the decoupling of their electronic states from the substrate (Figure 4).

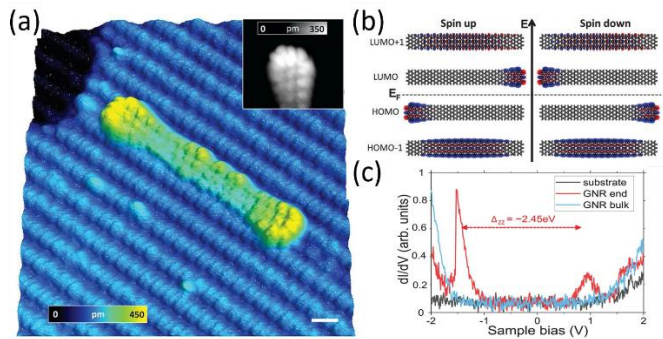

Figure 4. (a) STM image of atomically precise armchair GNR on rutile $TiO_2$ surface. (b) Nanoribbons have zigzag ends with coupled spin centers. (c) The insulating substrate enables observation of magnetic end states with a 2.45 eV splitting. (Source: Kolmer et al.)

STM measurements confirmed the existence of magnetic zigzag end states spatially separated by semiconducting nanoribbons. We observed the theoretically predicted spin-1/2 end-states with an exchange gap of $2.45 \pm 0.10$ eV (Figure 4c). Electronic states at the ends of such a narrow nanoribbon thus give rise to a pair of nonlocally entangled spins. When combined with recent developments in microwave transmitting scanning probe microscopy techniques, these electronically free-standing GNRs could enable qubit initialization, logic gate operation, and readout. STM manipulation of these nanoribbons and probing their transport properties with multiprobe STM can lead to the creation and testing of complex quantum circuits.

## CONCLUSIONS

Advanced, atomic-scale imaging, spectroscopy, and manipulation capabilities of STM make it possible to assemble multiple atoms into a chosen geometry so that defects can be created at will. STM thus offers a transformative approach to tame defects for QIS applications. Combining STM manipulation with in-situ synthesis and characterization capabilities, we are able to create atomic defects and interfaces that deliver novel quantum states, such as entangled spins, single-photon emissions, and atomic memories. These quantum defects, enable the creation of emerging qubits and memristors based on graphene, TMDs, and other 2D materials. Our research is focused on determining the host lattice–defect pairs, identifying suitable states for hosting quantum properties, creating these defects in a controllable manner, experimentally verifying the properties of these systems, and uncovering the fundamental physics responsible for these properties.

## IMPACT

Atomic defects provide a system that combines the environmental isolation necessary to maintain the coherence of quantum states with the ability to access and manipulate the states via electrical, magnetic, and optical probes. With proper utilization of these defects, not only can we create quantum states deterministically, but we can also control interactions and conversions of these states. This capability paves the way for QIS applications such as new ultrasensitive sensors in which quantum systems' extreme sensitivity to external perturbations is utilized, or quantum computing where quantum entangled and coherent ensembles are used for the physical representation and processing of information to tackle problems divergently complex for classic computers.

## COLLABORATE WITH ME

Saban M. Hus, hussm@ornl.gov

An-Ping Li, apli@ornl.gov

## ACKNOWLEDGMENTS

This work was supported by Center for Nanophase Materials Sciences (CNMS), which is a US Department of Energy, Office of Science User Facility at Oak Ridge National Laboratory.